\documentclass[twocolumn, aps,pre,superscriptaddress,showpacs,floatfix]{revtex4-2}

\UseRawInputEncoding
\usepackage{float}
\usepackage{dcolumn}
\usepackage{bm}
\usepackage[colorlinks = true, linkcolor = blue, urlcolor  = blue, citecolor = blue, anchorcolor = blue]{hyperref}
\usepackage{longtable}
\usepackage{booktabs}
\usepackage{mathrsfs}
\usepackage{graphicx,epsfig,latexsym,amssymb}
\usepackage{multirow,amsmath,array,booktabs,color}
\usepackage[section]{placeins}
\usepackage{color}
\usepackage{soul}

\begin{document}
\title{Ground state properties of the one-dimensional axial next-nearest-neighbor Ising model in a transverse field}
\author{Yun-Tong Yang}
\affiliation{School of Physical Science and Technology, Lanzhou University, Lanzhou 730000, China}
\affiliation{Lanzhou Center for Theoretical Physics, Key Laboratory of Theoretical Physics of Gansu Province, Key Laboratory of Quantum Theory and Applications of MoE, Gansu Provincial Research Center for Basic Disciplines of Quantum Physics, Lanzhou University, Lanzhou 730000, China}
\author{Fu-Zhou Chen}
\affiliation{School of Physical Science and Technology, Lanzhou University, Lanzhou 730000, China}
\affiliation{Lanzhou Center for Theoretical Physics, Key Laboratory of Theoretical Physics of Gansu Province, Key Laboratory of Quantum Theory and Applications of MoE, Gansu Provincial Research Center for Basic Disciplines of Quantum Physics, Lanzhou University, Lanzhou 730000, China}
\author{Hong-Gang Luo}
\email{luohg@lzu.edu.cn}
\affiliation{School of Physical Science and Technology, Lanzhou University, Lanzhou 730000, China}
\affiliation{Lanzhou Center for Theoretical Physics, Key Laboratory of Theoretical Physics of Gansu Province, Key Laboratory of Quantum Theory and Applications of MoE, Gansu Provincial Research Center for Basic Disciplines of Quantum Physics, Lanzhou University, Lanzhou 730000, China}

\begin{abstract}
Describing and understanding the consequences of competing interactions remains profoundly challenging in both classical and quantum systems, as it is difficult to identify suitable order parameters, thereby hindering the characterization of certain phases such as the floating phase found in the one-dimensional axial next-nearest-neighbor Ising (ANNNI) model in the presence of the frustration interactions between the nearest and next-nearest neighbor sites. In this work, we employ the pattern picture to explore the frustration physics in such a model. This picture has been comprehensively detailed in our previous article [Yang and Luo, Phys. Rev. E  \textbf{112}, 044102 (2025)]. Here, we apply it to the ANNNI model with periodic boundary conditions, considering system sizes ranging from $L=16$ to $L=128$. Our results demonstrate that the ground state of the system comprises four phases: ferromagnetic, paramagnetic, floating, and $\langle 2,2 \rangle$ antiphase. The transition from the ferromagnetic to the paramagnetic phase is continuous, analogous to that in the transverse-field Ising model, while the transition from the floating phase to the antiphase is first order. Furthermore, the floating phase exhibits particularly intriguing characteristics: states with distinct domain structures emerge successively as the frustration parameter increases. As the system size grows, this succession becomes progressively denser, leading to the reasonable inference that it eventually approaches a continuous variation in the thermodynamic limit. To validate the effectiveness of our picture, we computed the second derivative of the ground-state energy, which exhibits multiple dips within the floating phase$-$consistent with the pattern language. 
\end{abstract}

\pacs{}
\maketitle

\section{\label{sec:level1}Introduction}
In recent decades, frustrated spin systems have become a highly prominent topic in the field of quantum magnetism due to the emergence of various exotic physical states \cite{Lacroix2011, Diep2013, Starykh2015, Vojta2018}, such as quantum spin liquids \cite{Anderson1973, Kitaev2006, Balents2010, Savary2017, Zhou2017, Wen2019, Broholm2020}. At the same time, this remains an extremely challenging subject, as reliable computational methods are still lacking, for examples, the size limitations of exact diagonalization \cite{Lin1990} and the sign problem in Monte Carlo simulations \cite{Assaad2008, Sandvik2010}. Furthermore, for a given frustrated model, the phase diagram including the locations of phase transitions and the nature of these transitions often depends on the methods employed, with different approaches yielding inconsistent results \cite{Nagy2011, Fumani2021, Bonfim2017}. This underscores the ongoing need to develop effective methods for studying frustrated systems.

In this article, we discuss the simplest frustrated model, namely, the one-dimensional axial next-nearest-neighbor Ising (ANNNI) model \cite{Elliott1961, Fisher1980, Bak1982, Selke1988}, which was named by Fisher and Selke \cite{Fisher1980} in 1980. The Hamiltonian of the model is given by: 
\begin{equation}
\hat{H} = - J_1\sum_{\langle i,j\rangle}\hat\sigma^z_i \hat\sigma^z_{j} + J_2\sum_{\langle\langle i,j\rangle\rangle}\hat\sigma^z_i \hat\sigma^z_{j}- h_x \sum_i \hat\sigma^x_i, \label{annni0}
\end{equation}
where frustration arises from the competition between the ferromagnetic interaction of the first term $(J_1>0)$ and the antiferromagnetic interaction of the second term $(J_2>0)$. When $h_x=0$, the model reduces to a classical one, undergoing a first-order phase transition from a ferromagnetic phase to a $\langle 2,2 \rangle$ antiphase at $J_2 = J_1/2$. When $h_x \neq 0$, quantum fluctuations occur due to the commutation relation $\left[\sigma^z,\sigma^x \right]=2i\sigma^y$, significantly enriching the ground-state properties of the ANNNI model. Existing studies have reported divergent results, which are briefly summarized as follows:

(i) The ground state consists of five phases: ferromagnetic, unmodulated paramagnetic, modulated paramagnetic, floating, and antiphase. These results were obtained through quantum Monte Carlo \cite{Arizmendi1991}, exact diagonalization of small lattice systems \cite{Sen1992, Fumani2021}, density matrix renormalization group \cite{Beccaria2006, Beccaria2007}, and the Jordan-Wigner transformation \cite{Fumani2021}.

(ii) The ground state comprises four phases: ferromagnetic, paramagnetic, floating, and antiphase. These findings were derived from exact diagonalization \cite{Rieger1996}, perturbation theory \cite{Chandra2007}, the matrix product state method \cite{Nagy2011}, the Variational Quantum Eigensolver solutions \cite{Lively2024}, and a feedback-based quantum algorithm \cite{Pexe2024}.

(iii) The ground state exhibits three phases: ferromagnetic, paramagnetic, and antiphase. These were identified using the interface approach \cite{Sen1997},  the finite-size scaling \cite{Guimaraes2002}, quantum conventional neural networks \cite{Monaco2023}, and machine learning \cite{Ferreira2024}.

(iv) The ground state possesses an infinite number of phases: quantum fidelity methods confirm the existence of ferromagnetic, floating, and antiphase phases, and reveal an infinite number of modulated phases between the ferromagnetic and floating phases \cite{Bonfim2017, Bonfim2020}.

These phases are defined through two-site correlation functions \cite{Sen1992, Rieger1996, Beccaria2006, Beccaria2007, Chandra2007}. The observed inconsistencies indicate that, even for the simplest frustrated model, determining phases solely by computing correlation functions is insufficient. This current dilemma arises partly from the inherent difficulty in identifying order parameters within frustrated systems. However, quantum mechanics teaches us that once the wave function of a system is obtained, all information about the system is, in principle, accessible. Therefore, the present situation also stems from the lack of effective methods to extract relevant physical information from the wave function.

Here, we employ the pattern language to characterize the ground state properties of the ANNNI model. This approach has previously been applied to the quantum Rabi model \cite{Yang2024} and the transverse-field Ising model \cite{Yang2025b}, describing the superradiant phase transition and the paramagnetic-ferromagnetic phase transition, respectively. The core idea of this method is to derive the collective modes of the system from the operator space as ``patterns", use these patterns to decompose the Hamiltonian into multiple sub-Hamiltonians, and then project the wave function onto these sub-Hamiltonians to observe their evolution with varying model parameters. For the ANNNI model, the floating phase is the most challenging to describe, yet the pattern language clearly reveals the physical processes occurring in this phase. As shown by the results in Sec. \ref{sec:level2} and \ref{sec:level3}, patterns function similarly to order parameters or correlation functions, making them particularly convenient for characterizing phases of the system. Thus, our method provides an effective approach for extracting physical information from the wave function.

The structure of this paper is as follows. In Sec. \ref{sec:level2}, we introduce the pattern language, illustrating its application through a classical model and a small-scale quantum model with a chain length of $L=8$. In Sec. \ref{sec:level3}, we combine the density matrix renormalization group (DMRG) with the pattern language to perform calculations on larger systems, thereby demonstrating the ground-state properties$-$particularly those of the floating phase. Sec. \ref{sec:level4} is devoted to a conclusion.

\section{\label{sec:level2} The pattern language}
The pattern language requires performing two successive diagonalizations of the Hamiltonian. The purpose of the first diagonalization is to obtain all the patterns of the system considered in an operator space, while the second diagonalization performed in a basis space aims to derive the wave function of the system. In this section, we demonstrate the specific procedure of this method using a classical model and a quantum model with a lattice size of $L = 8$. Although both examples are relatively trivial since small lattice size, they serve as suitable cases for illustrating the methodology.

\subsection{\label{sec:level2A} The classical case}
\begin{figure}[tbp]
\begin{center}
\includegraphics[width = 0.9\columnwidth]{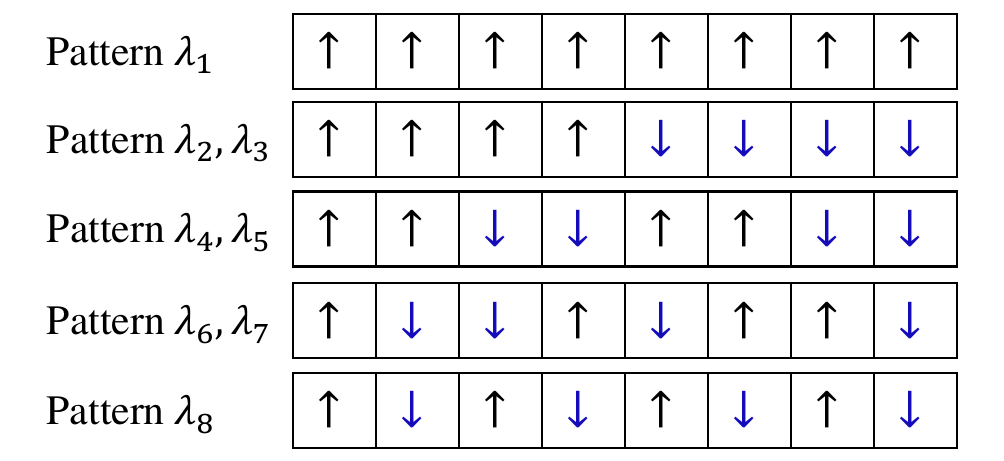}
\caption{Patterns of the Hamiltonian in Eq. (\ref{annni1}). Spin configurations (up/down) correspond to the signs of the coefficients $u_{n,i}$ in $A_n$. Results are shown for a system with lattice size $L=8$ under PBC. These patterns are roubst and remain unchanged across different parameters of the model.}\label{fig1}
\end{center}
\end{figure}

We first consider the case where $h_x=0$ in the Eq. (\ref{annni0}), namely, the classical frustrated model,
\begin{equation}
H = - J \sum_i \sigma^z_i \sigma^z_{i+1} + \kappa J \sum_i \sigma^z_i \sigma^z_{i+2}. \label{annni1}
\end{equation}
where $\kappa$ is a dimensionless frustration parameter. For the chain length $L$ with periodic boundary condition (PBC), this Hamiltonian can be rewritten as
\begin{eqnarray}
H &=& \left(
\begin{array}{ccccc}
\sigma^z_1 & \sigma^z_2 & \sigma^z_{3} & \cdots & \sigma^z_L
\end{array}
\right)\nonumber\\
&\times&
\left(
\begin{array}{ccccc}
0 & -J & \kappa J & \cdots & -J \\
-J & 0 & -J &\cdots & \kappa J \\
\kappa J &-J &0 &\cdots &0 \\
\vdots &\vdots &\vdots &\ddots &\vdots \\
- J & \kappa J &0 & \cdots & 0 
\end{array}
\right)\nonumber\\
&&\times \left(
\begin{array}{ccccc}
\sigma^z_1 & \sigma^z_2 & \sigma^z_{3} & \cdots & \sigma^z_L
\end{array}
\right)^T,\label{annni2}
\end{eqnarray}
The matrix in Eq. (\ref{annni2}) has dimensions $L\times L$. Diagonalization yields eigenvalues and corresponding eigenfunctions $\{\lambda_n, u_n\} (n = 1, 2, \cdots, L)$. Using these results, the Hamiltonian can be rewritten as 
\begin{equation}
H = \sum_{n=1}^{L} H_n = \sum_{n=1}^{L} \lambda_n A_n^2, \label{annni3a} 
\end{equation}
where $A_n$ is the linear combination of lattice spin $\sigma^z_i$, which reads 
\begin{equation}
A_n = \sum_{i=1}^{L}u_{n,i}\sigma^z_{i}, \label{annni3b}
\end{equation}
where $u_{n,i}$ is the corresponding wave vector. We first discuss the properties of $A_n$ using $L=8$ as an example. After diagonalizing the matrix in Eq. (\ref{annni2}), we extract the signs of the coefficients $u_{n,i}$ in $A_n$, mapping negative signs to spin-up and positive signs to spin-down, which yields Fig. \ref{fig1}. This figure represents the collective modes of all lattice sites induced by interactions, exhibiting a domain structure. We refer to these as ``patterns" and denote them by $\lambda_n$'s. It can be observed that pattern $\lambda_1$ contains only a single domain, corresponding to a ferromagnetic configuration; patterns $\lambda_{2,3}$ consist of two domains; patterns $\lambda_{4,5}$ consist of four domains; patterns $\lambda_{6,7}$ consist of six domains; and the final pattern $\lambda_{8}$ possesses eight domains, corresponding to an antiferromagnetic configuration. As the system size increases, the number of patterns increases accordingly, evolving from the single-domain pattern $\lambda_1$ to the $L$-domain pattern $\lambda_L$.

\begin{figure}[tbp]
\begin{center}
\includegraphics[width = 0.9\columnwidth]{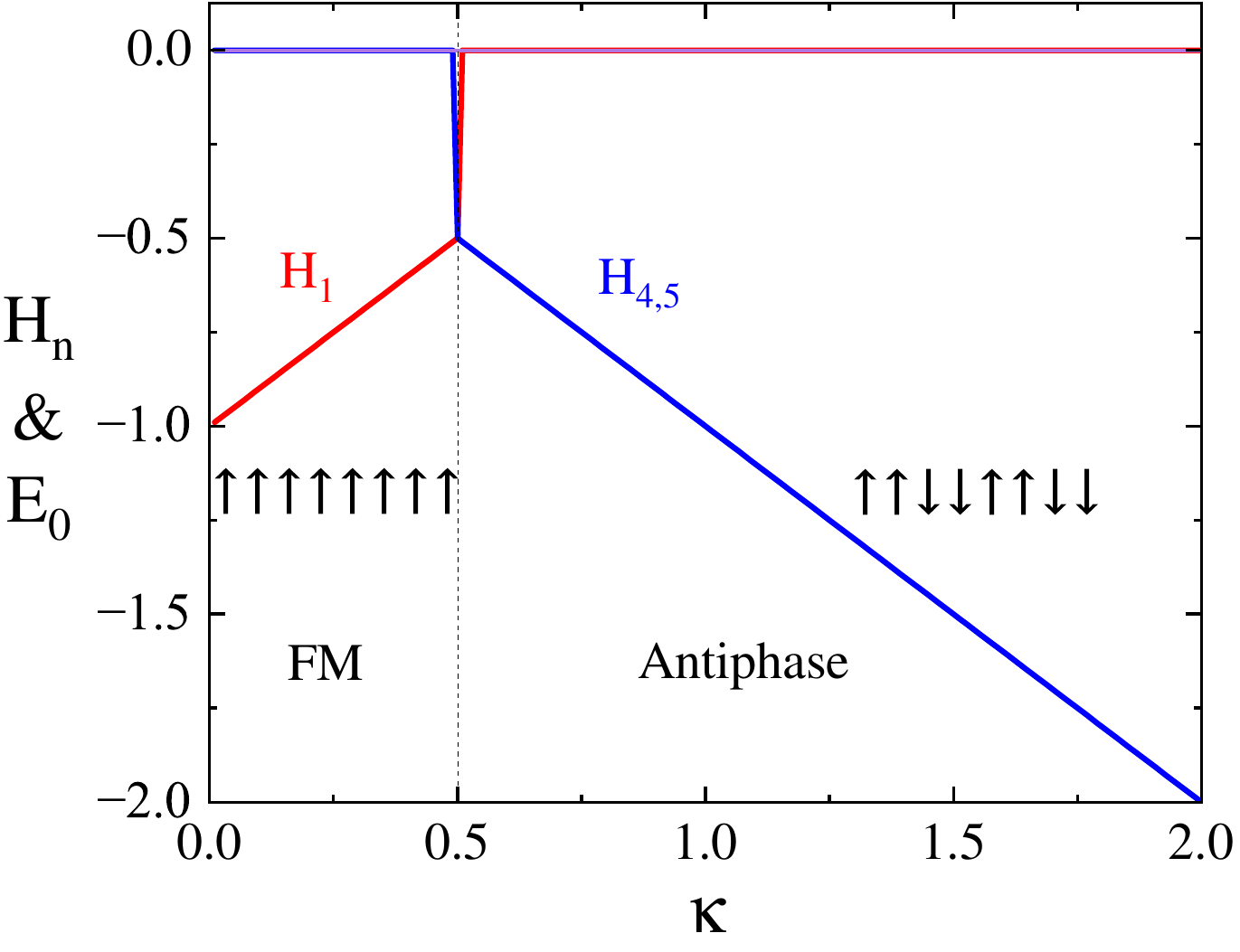}
\caption{The total energy of the system (not shown) and the energies of the sub-Hamiltonians (colored curves) as functions of $\kappa$. The red curve corresponds to $H_1$, while the blue curve represents $H_{4,5}$. Only the sub-Hamiltonian energies are displayed, since the energies of $H_1$ and $H_{4,5}$ coincide with the total energy. This indicates that the system is in the ferromagnetic phase at $\kappa<0.5$ and in the antiphase at $\kappa>0.5$. All other sub-Hamiltonians have zero energy.}\label{fig2}
\end{center}
\end{figure}

After obtaining these patterns, we calculate the total energy of the system and the energies of each sub-Hamiltonian $H_n$, and examine the weight of each sub-Hamiltonian's energy contribution to the total energy to determine the state of the system. As shown in Fig. \ref{fig2}, for $\kappa < 0.5$, $H_1$ contributes the entire energy of the system. Since $H_1$ corresponds to pattern $\lambda_1$, the system is in the ferromagnetic phase. At $\kappa = 0.5$, the energy of $H_1$ drops abruptly to zero, while the energies of $H_{4,5}$ jump from zero to finite values. For $\kappa > 0.5$, $H_{4,5}$ contribute fully to the system's energy. Given that $H_{4,5}$ correspond to patterns $\lambda_{4,5}$, the system is in the $\langle 2,2 \rangle$ antiphase.

Although this is a relatively trivial and exactly solvable example, it demonstrates that the proposed patterns effectively serve as order parameters in the classical model. Therefore, our framework provides a natural approach for identifying order parameters. In the following, we apply this framework to the quantum model. 

\subsection{\label{sec:level2B} The quantum case}
\begin{figure}[tbp]
\begin{center}
\includegraphics[width = \columnwidth]{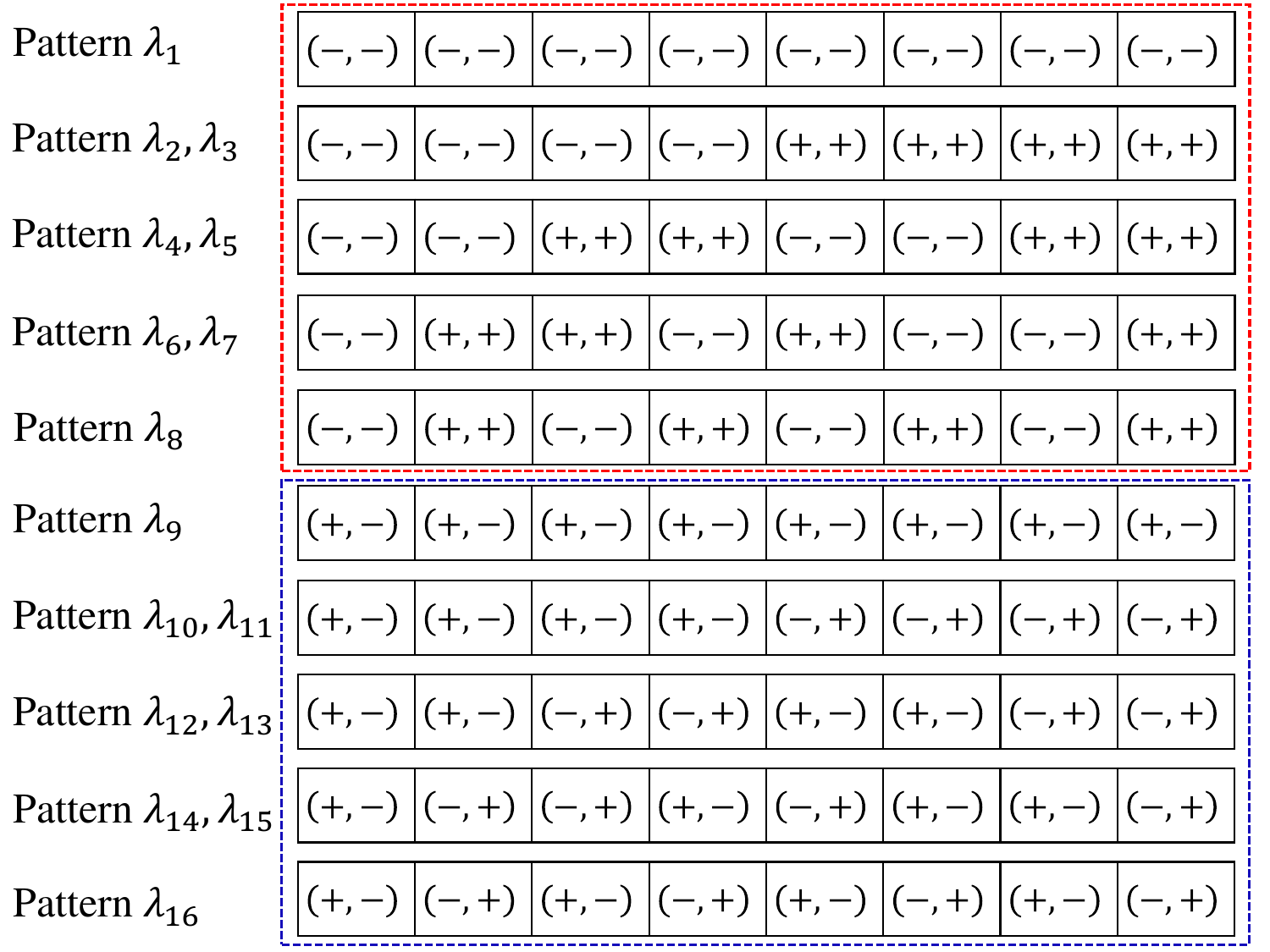}
\caption{Patterns and their relative phases obtained from the first diagonalization. Each pattern is labeled by the single-body operators $\hat{A}_n$ where the notation $(\pm,\pm)$ denotes the signs of the coefficients $(u_{n,2i-1},u_{n,2i})$. The patterns are divided into two groups, marked by the red and blue dashed frames, corresponding to $\lambda_n < 0$ and $\lambda_n > 0$, respectively. Results are shown for a system size of $L=8$. These patterns remain unchanged for different parameters of the model.}\label{fig3}
\end{center}
\end{figure}

Now we discuss the case where $h_x\neq 0$ in the Eq. (\ref{annni0}). In the following we take $h_x$ as units of energy. The Hamiltonian reads
\begin{equation}
\hat{H} = - J\sum_{i}\hat\sigma^z_i \hat\sigma^z_{i+1} + \kappa J\sum_{i}\hat\sigma^z_i \hat\sigma^z_{i+2}- \sum_i \hat\sigma^x_i. \label{annni4}
\end{equation}
This Hamiltonian can be reformulated as 
\begin{eqnarray}
\hat{H} &=& \left(
\begin{array}{ccccccccc}
 -i\hat{\sigma}^y_1& \hat{\sigma}^z_1& -i\hat{\sigma}^y_2& \hat{\sigma}^z_2& -i\hat{\sigma}^y_{3}& \hat{\sigma}^z_{3} & \cdots & -i\hat{\sigma}^y_L& \hat{\sigma}^z_L
\end{array}
\right)\nonumber\\
&\times&
\left(
\begin{array}{ccccccccc}
0 &-1 &0 &0 & 0 &0 &\cdots &0 &0 \\
-1 &0 &0 &- J & 0&\kappa J &\cdots &0 &- J \\
0 &0 &0 &-1 &0&0 &\cdots &0 &0 \\
0 &- J &-1 &0 &0&0 &\cdots &0 &\kappa J \\
0 &0 &0 &0 &0 &0 &\cdots &0 &0 \\
0 &\kappa J &0 &0 &0 &0 &\cdots &0 &0 \\
\vdots &\vdots &\vdots &\vdots &\vdots &\vdots &\ddots &\vdots &\vdots \\
0 &0 &0 &0 &0&0&\cdots &0 &-1 \\
0 &- J &0 &\kappa J &0&0 &\cdots &-1 &0 
\end{array}
\right)\nonumber\\
&&\times \left(
\begin{array}{ccccccccc}
 i\hat{\sigma}^y_1& \hat{\sigma}^z_1& i\hat{\sigma}^y_2& \hat{\sigma}^z_2& i\hat{\sigma}^y_{3}& \hat{\sigma}^z_{3} & \cdots & i\hat{\sigma}^y_L& \hat{\sigma}^z_L
\end{array}
\right)^T,\label{annni5}
\end{eqnarray}
where the identity of Pauli matrices $\hat{\sigma}^y \hat{\sigma}^z = i \hat{\sigma}^x$ has been used to the transverse-field term. The dimension of this matrix is $2L\times 2L$. Using its eigenvalues and eigenfunctions $\{\lambda_n, u_n\} (n = 1, 2, \cdots, 2L)$, the Hamiltonian is rewritten as 
\begin{equation}
\hat{H} = \sum_{n=1}^{2L} \hat{H}_n = \sum_{n=1}^{2L} \lambda_n \hat{A}^\dagger_n \hat{A}_n, \label{annni6a} 
\end{equation}
where the operator $\hat{A}_n$ composes of single-body operators
\begin{equation}
\hat{A}_n = \sum_{i=1}^{L}\left[u_{n,2i-1} (i\hat{\sigma}^y_{i}) + u_{n,2i} \hat{\sigma}^z_{i}\right]. \label{annni6b}
\end{equation}
This completes the first diagonalization of the Hamiltonian in the operator space. We again begin by examining the properties of $\hat{A}_n$ for the case of $L=8$. By extracting the signs of $u_{n,2i-1}$ and $u_{n,2i}$, we obtain Fig. \ref{fig3}, where each pair of parentheses represents a lattice site. Here, $u_{n,2i}$ corresponds to $u_{n,i}$ in the classical case, and its sign can be interpreted as spin up or spin down. However, since each site is now associated with two operators, a simple spin-up/spin-down representation is insufficient. Therefore, we directly use positive and negative signs for representation. As shown in Fig. \ref{fig3}, pattern $\lambda_1$ contains only a single domain, corresponding to a ferromagnetic configuration; patterns $\lambda_{2,3}$ exhibit two domains; patterns $\lambda_{4,5}$ exhibit four domains; patterns $\lambda_{6,7}$ exhibit six domains; and finally, pattern $\lambda_8$ have eight domains, corresponding to an antiferromagnetic configuration. This is fully consistent with the classical case. The difference lies in the quantum case, where the number of patterns doubles, with additional patterns from $\lambda_9$ to $\lambda_{16}$, whose domain structures also evolve from ferromagnetic to antiferromagnetic. However, the two groups (highlighted by the red and blue frames in Fig. \ref{fig3}) differ in the signs of $u_{n,2i-1}$, reflecting the influence of quantum effects and distinguishing the quantum model from its classical counterpart.

\begin{figure}[tbp]
\begin{center}
\includegraphics[width = 0.9\columnwidth]{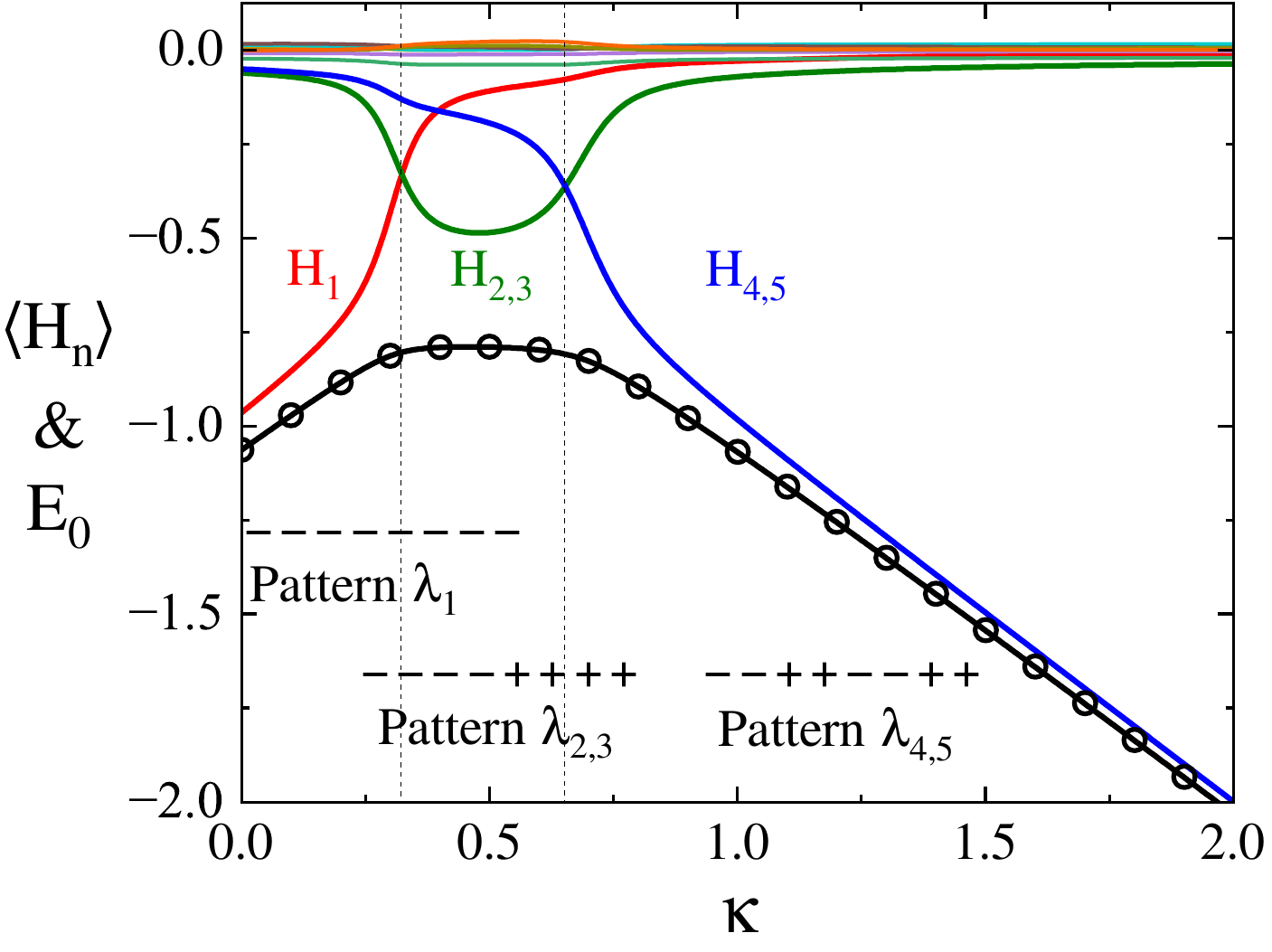}
\caption{Ground state energies as functions of the frustration parameter $\kappa$, computed using Eq. (\ref{annni6a}) (black solid lines) and direct numerical ED (circles). The corresponding energy components $\langle H_n \rangle$ are shown as colored solid lines. The patterns displayed below are obtained from the signs of the coefficients $u_{n,2i}$. The model parameter is taken as $J=2.0$.}\label{fig4}
\end{center}
\end{figure}

Now we insert the complete basis $|\{\sigma^z_i\}\rangle (i = 1, 2,\cdots, L)$ into the Hamiltonian (\ref{annni6a}), as done in Ref. \cite{Yang2025a, Yang2025b, Yang2025c}. The matrix form of operator $\hat{A}_n$ is $\left[\hat{A}_n\right]_{\{\sigma^z_i\},\{\sigma^z_i\}^{\prime}} = \langle\{\sigma^z_i\}|\hat{A}_n|\{\sigma^z_i\}^{\prime}\rangle$. Then the matrix of the Hamiltonian (\ref{annni6a}) is written as 
\begin{eqnarray}
&& \langle \{\sigma^z_i\}|\hat{H}|\{\sigma^z_i\}^{\prime}\rangle = \sum_{n=1}^{2L} \lambda_n \nonumber\\
&& \hspace{1cm}\times \sum_{\{\sigma^z_i\}^{\prime\prime}} \left[\hat{A}^\dagger_n\right]_{\{\sigma^z_i\},\{\sigma^z_i\}^{\prime\prime}}\left[\hat{A}_n\right]_{\{\sigma^z_i\}^{\prime\prime},\{\sigma^z_i\}^{\prime}}.\label{annni7}
\end{eqnarray}
This constitutes the second diagonalization of the Hamiltonian, aimed at obtaining the wave function of the model. This diagonalization is largely similar to ED. In this work, we focus exclusively on the ground-state properties, with the ground-state wave function denoted as $\Psi_0$. By projecting $\Psi_0$ onto each sub-Hamiltonian, i.e. $\langle H_n \rangle=\langle \Psi_0|\hat{H}_n|\Psi_0\rangle$, we obtain the results shown in Fig. \ref{fig4}. The model parameters here are $L=8, J=2$ with varying $\kappa$. The black curve represents the total energy, while the colored curves correspond to the decomposed energies, whose sum equals the black curve. The circles indicate results from direct ED, which are in full agreement with those from our method. At small $\kappa$, $H_1$ contributes predominantly to the total energy. As $\kappa$ increases, the weight of $H_1$ decreases, and $H_{2,3}$ become the main contributors. With further increase in $\kappa$, $H_{4,5}$ dominate. This relay-like energy change is a direct manifestation of frustration effects. By identifying which sub-Hamiltonian dominates in different parameter regions, we determine which pattern is active and thus infer the state of the system. In this sense, the patterns function as order parameters. To characterize the phase diagram of the ground state, $L=8$ is clearly insufficient. In Sec. \ref{sec:level3}, we extend the calculations to larger system sizes, following the same methodology as for $L=8$.

\section{\label{sec:level3} Ground state properties of the ANNNI model} 
In this section, we employ the DMRG method to compute the wave functions of the Hamiltonian for system sizes ranging from $L=16$ to $L=128$. We examine three cases: $J=2, J=1$ and $J=0.5$, which at $\kappa=0$ correspond to the ferromagnetic phase, the critical point, and the paramagnetic phase, respectively. Our focus is on elucidating the physical nature of the floating phase. 

\subsection{\label{sec:level3A} The $J=2.0$ case}
\begin{figure}[tbp]
\begin{center}
\includegraphics[width = \columnwidth]{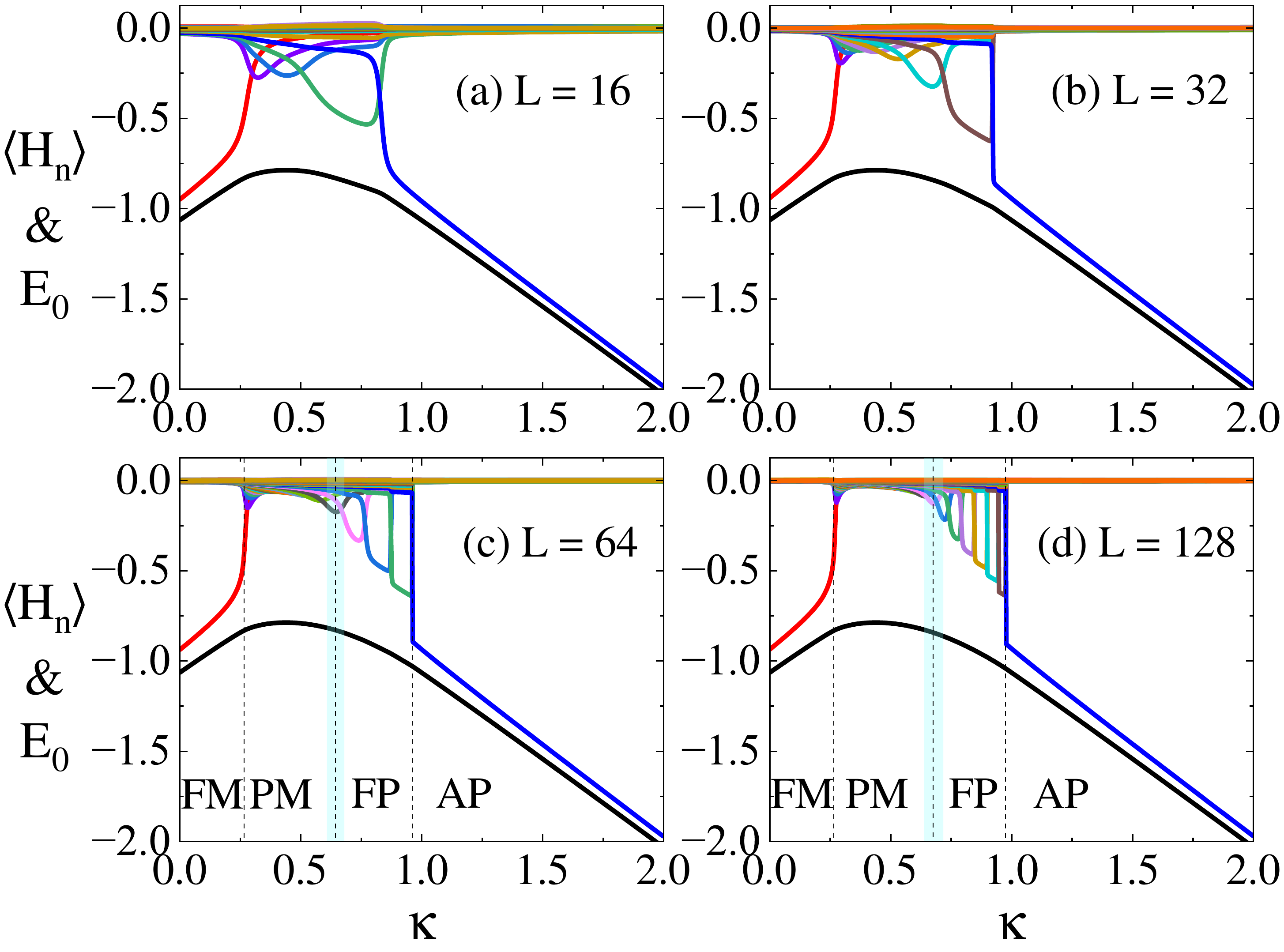}
\caption{Ground state energy $E_0$ (black ines) and the energy components $\langle H_n \rangle$ (colored lines) as functions of the frustration parameter $\kappa$ for system sizes ranging from $L=16$ to $128$. Four distinct phases are identified: ferromagnetic (FM), paramagnetic (PM), floating phase (FP) and antiphase (AP). Vertical dashed lines indicate phase boundaries with shaded region denoting a crossover. The model parameter is fixed at $J=2.0$. The step of $\kappa$ is set to 0.003 for this and subsequent analyses.}\label{fig5}
\end{center}
\end{figure}

Figure \ref{fig5} shows the ground-state energy (black curve) and the energies of the decomposed sub-Hamiltonians (colored curves) for systems of different sizes. For $L=16$ and $32$, at small $\kappa$, the red curve contributes predominantly to the total energy, corresponding to pattern $\lambda_1$ with a ferromagnetic configuration. At large $\kappa$, the blue curve dominates, corresponding to patterns $\lambda_{\frac{L}{2},\frac{L}{2}+1}$ with a $\langle 2,2 \rangle$ spin configuration. In the intermediate $\kappa$ range, multiple $H_n$ emerge successively, however, the physical state of the system remains unclear due to the small system sizes. When the system size is increased to $L=64$, the small- and large-$\kappa$ regions still correspond to ferromagnetic and $\langle 2,2 \rangle$ configuration, respectively, while the intermediate region becomes clearer. In the approximate range $\kappa\in(0.25,0.7)$, all sub-Hamiltonians contribute nearly equally to the energy, indicating the absence of a dominant order$-$a signature of a disordered (paramagnetic) phase. As $\kappa$ increases further, the system enters a new regime where adjacent $H_n$ with different domain structures emerge successively and evolve rapidly with $\kappa$, showing an increasing number of domains until the system settles into the $\langle 2,2 \rangle$ configuration represented by the blue curve. This behavior becomes more pronounced at $L=128$, as shown in Fig. \ref{fig5}(d). In the region $\kappa\in(0.7,1.0)$, the system undergoes rapid changes with varying $\kappa$, a characteristic of the floating phase.

Based on the behavior of the sub-Hamiltonians, the ground state can be divided into four regions: ferromagnetic phase, paramagnetic phase, floating phase, and $\langle 2,2 \rangle$ antiphase. The transition from the ferromagnetic to the paramagnetic phase is continuous, analogous to that in the transverse-field Ising model. The transition from the floating phase to the $\langle 2,2 \rangle$ antiphase is first-order, as indicated by the sharp change in blue curve. In contrast, the boundary between the paramagnetic and floating phases is not sharply defined and resembles a crossover, which we indicate using a shaded region.

\begin{figure}[tbp]
\begin{center}
\includegraphics[width = \columnwidth]{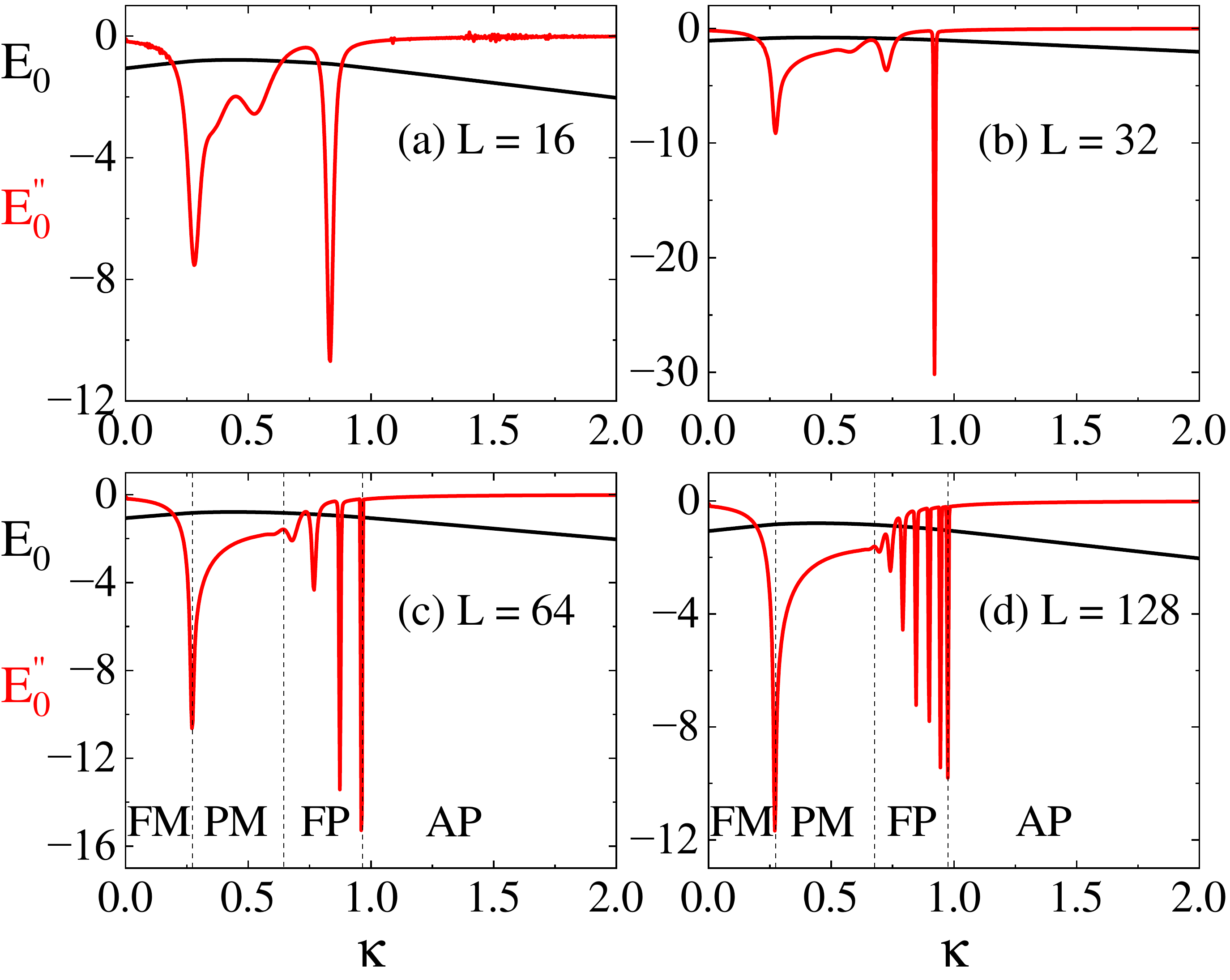}
\caption{Ground state energy $E_0$ (black ines) and its corresponding second derivative (red lines) for different system sizes. Four distinct phases are identified, consistent with Fig. \ref{fig5}. Phase boundaries are marked by vertical dashed lines. The model parameter is fixed at $J=2.0$. }\label{fig6}
\end{center}
\end{figure}

The above analysis can also be verified by examining the second derivative of the energy, as shown in Fig. \ref{fig6}, where the black curve represents the total energy and the red curve corresponds to its second derivative. In the floating phase, the second derivative exhibits multiple dips, consistent with the results obtained from our pattern language framework. However, the pattern language further reveals the internal state of the system within the floating phase. It can also be observed from Fig. \ref{fig6} that the second derivative varies smoothly across the transition from the paramagnetic to the floating phase. This transition is conjectured to be of infinite order, known as the Kosterlitz-Thouless transition in the literatures \cite{Kosterlitz1972, Beccaria2006, Nagy2011, Fumani2021, Ferreira2024, Cea2024}. From our picture, its behaviour resembles a crossover, where frustration drives the system to gradually evolve from a disordered state into a regime characterized by relay-like energy transfer.

\begin{figure}[tbp]
\begin{center}
\includegraphics[width = \columnwidth]{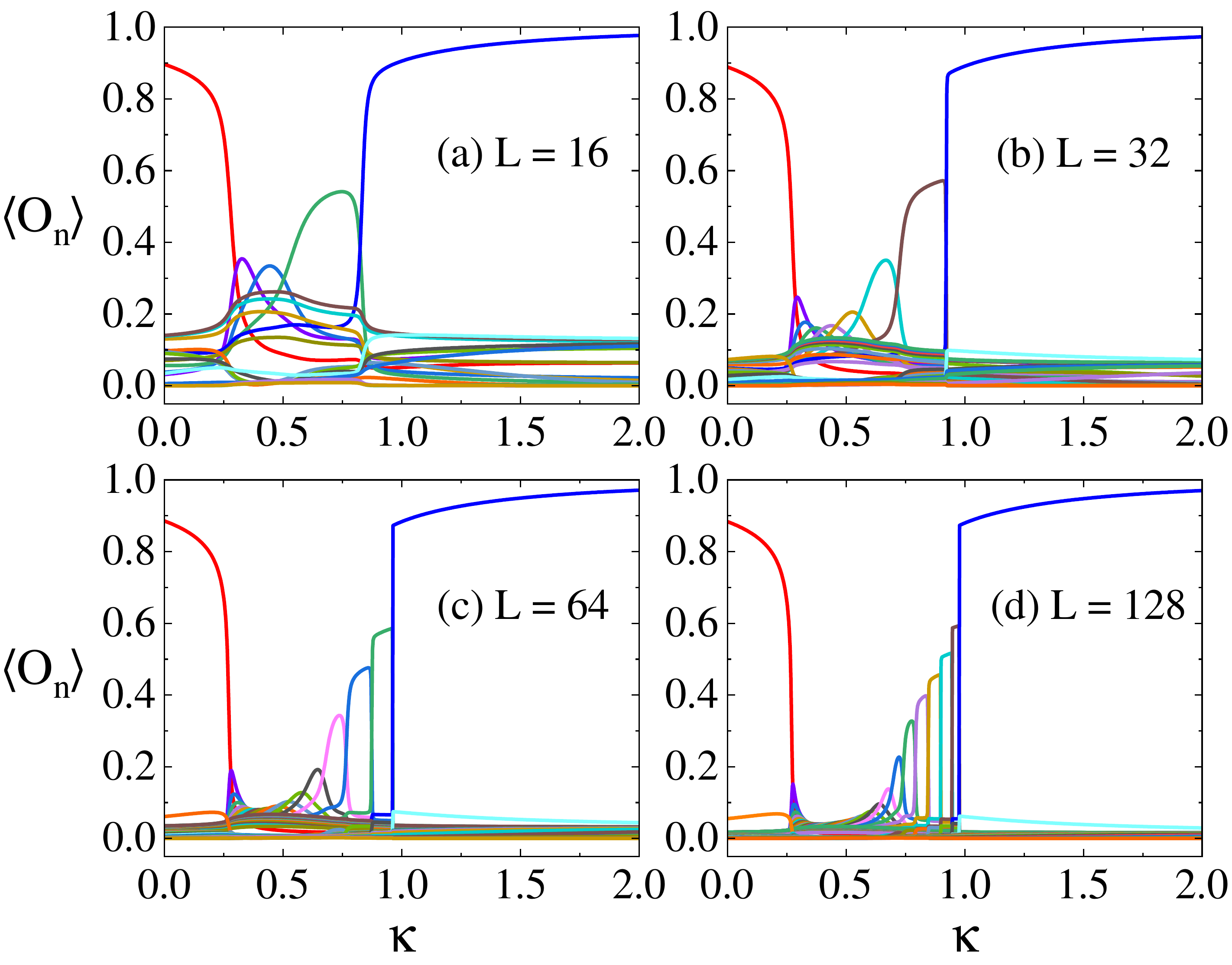}
\caption{Ground-state pattern occupancy as functions of $\kappa$ for system sizes ranging from $L=16$ to $128$. The model parameter is fixed at $J=2.0$.}\label{fig7}
\end{center}
\end{figure}

In addition to the ground-state energy of the system, we define a new physical quantity using the operator $\hat{A}_n$, termed the pattern occupancy:
\begin{equation}
\langle O_n \rangle = \langle\Psi_0|\hat{A}^\dagger_n \hat{A}_n|\Psi_0\rangle. \label{occ0}
\end{equation}
The results are shown in Fig. \ref{fig7}. It can be observed that all the preceding analyses are reflected in this quantity, indicating that it also comprehensively captures the physical information of the system. It is worth noting that the values of this quantity are always positive, and its behavior closely resembles that of an order parameter. Furthermore, the positive definiteness of this operator makes it suitable for Monte Carlo simulations, potentially offering a pathway to mitigate the sign problem. This represents one of the directions we intend to explore in future work.

\subsection{\label{sec:level3B} The $J=1.0$ and $J=0.5$ cases}
\begin{figure}[tbp]
\begin{center}
\includegraphics[width = \columnwidth]{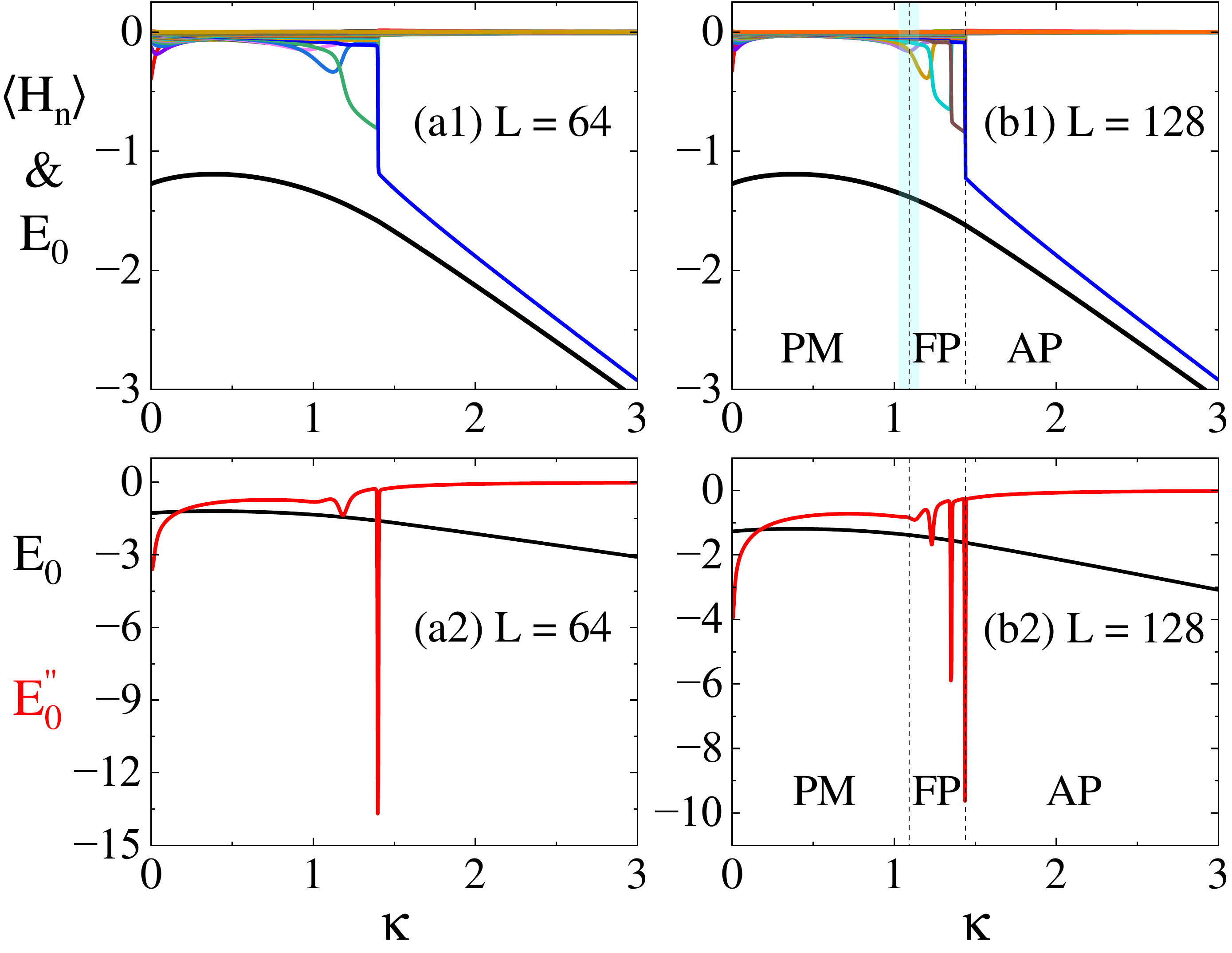}
\caption{(a1) $\&$ (b1) Ground state energy $E_0$ (black ines) and energy components $\langle H_n \rangle$ (colored lines) as functions of $\kappa$. (a2) $\&$ (b2) Ground state energy $E_0$ (black ines) and its corresponding second derivative (red lines). System sizes are $L=64$ and $128$, respectively. Three phases are identified: paramagnetic (PM), floating phase (FP) and antiphase (AP). Vertical dashed lines indicate phase boundaries, with the shaded region representing a crossover. The model parameter is fixed at $J=1.0$. }\label{fig8}
\end{center}
\end{figure}

\begin{figure}[tbp]
\begin{center}
\includegraphics[width = \columnwidth]{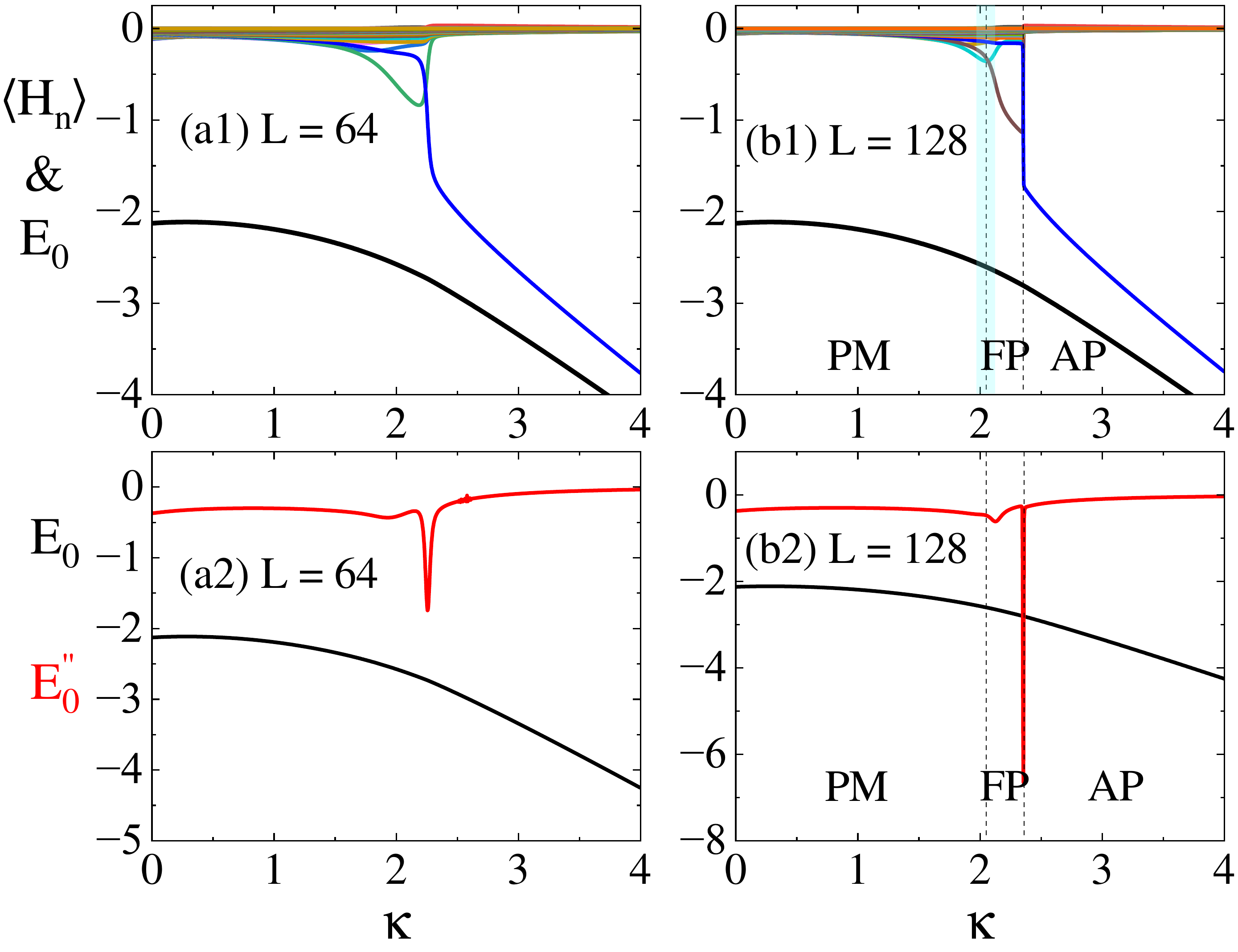}
\caption{All specifications remain identical to those in Fig. \ref{fig8}, except that the model parameter is fixed at $J=0.5$.}\label{fig9}
\end{center}
\end{figure}

We now discuss the cases of $J=1$ and $J=0.5$, which at $\kappa=0$ correspond to the critical point and the paramagnetic phase of the transverse-field Ising model, respectively. As shown in Fig. \ref{fig8} and Fig. \ref{fig9}, no ferromagnetic phase appears as $\kappa$ varies. With increasing $\kappa$, the system transitions from the paramagnetic phase to the floating phase and then to the $\langle 2,2 \rangle$ antiphase. Compared to the case of $J=2$, the paramagnetic phase region is enlarged, while the floating phase region is reduced. These results are consistent with the discussion in Sec. \ref{sec:level3} A.

\section{\label{sec:level4}Conclusion}
We have investigated the ground-state properties of the ANNNI model using the pattern language approach. Our results demonstrate that the ground state comprises four distinct phases: ferromagnetic, paramagnetic, floating, and $\langle 2,2 \rangle$ antiphase. Among these, the transition from the ferromagnetic to the paramagnetic phase is of second order, while that from the floating phase to the antiphase is first order. In particular, we have characterized the floating phase by the successive emergence of states with varying patterns or domain numbers, manifested as a relay-like energy change among sub-Hamiltonians as the frustration parameter varies. As the lattice size tends to infinity, this relay-like behavior evolves into a continuous variation. This work presents the first theoretical visualization of the microscopic structure of the frustration-induced floating phase. 

Finally, we emphasize that the starting point of the pattern language is not based on individual particles, but rather on the collective modes arising from their organization. These collective modes effectively serve as the ``order parameters" of the system. Therefore, the essence of the pattern language lies not in describing the motion of individual particles, but in characterizing the relationships among them. From this perspective, the pattern language offers a novel pathway for characterizing phase transitions.

\section{Acknowledgments}
The work is partly supported by the National Key Research and Development Program of China (Grant No. 2022YFA1402704) and the programs for NSFC of China (Grant No. 12247101).

%

\end{document}